\documentclass[reprint,amsmath,amssymb,aps,prd]{revtex4-2}

\usepackage{dcolumn}
\usepackage{bm}
\usepackage{subcaption}
\usepackage{hyperref}
\usepackage[pdftex]{graphicx}
\usepackage{amsmath, amssymb, amsthm, braket,mathrsfs}
\usepackage{bbold}
\usepackage{bm}

\usepackage{tikz}
\usepackage{tikz-3dplot}
\usetikzlibrary{calc}
\usepackage[T1]{fontenc}

\DeclareMathOperator{\U}{U}

\begin{document}

\preprint{APS/123-QED}

\title{Baby skyrmion crystals stabilized by vector mesons}
\author{P.~Leask}\email{mmpnl@leeds.ac.uk}
\affiliation{School of Mathematics, University of Leeds, Leeds, LS2 9JT, England, UK}

\date{\today}

\begin{abstract}
In this letter we study soliton crystals in the $(2+1)$-dimensional analogue model of the $(3+1)$-dimensional Adkins–Nappi model of nuclear physics.
The baby $\omega$-Skyrme model studied here is an $O(3)$ nonlinear $\sigma$ model coupled to a massive vector meson, the $\omega$-meson.
Using recently developed methods in the $(3+1)$-dimensional $\omega$-Skyrme model we are able to construct soliton crystals in this $(2+1)$-dimensional baby $\omega$-Skyrme model.
The resulting crystals form a hexagonal lattice structure and are qualitatively \emph{and} quantitatively similar to crystals observed in the standard baby Skyrme model.
\end{abstract}

\maketitle



\section{Introduction}

It is well known that the phase structure of nuclear matter is rich and highly non-trivial.
At high densities, the hadrons have considerably different properties than in the lower density regimes.
In order to understand what happens to nuclear matter under extreme conditions, the underlying theory must be consistent with quantum chromodynamics (QCD).
A detailed analysis by 't Hooft \cite{Hooft_1974} showed that, in the large $N_c$-limit, low-energy QCD can be reduced to an effective chiral field theory of mesons.
Witten \cite{Witten_1979} took this further and conjectured that baryons arise as solitons in this large-$N_c$ theory.

Skyrme's original model \cite{Skyrme_1961} is one such description; it is an effective Lagrangian involving only the lightest of mesons, the pions, with the idea that baryons emerge as stable solitons with non-trivial topological charge.
At its core, the Skyrme model contains the nonlinear $\sigma$ model (NL$\sigma$M).
By simple application of Derrick's Theorem, the solitons are not energetically stable as the NL$\sigma$M is not length scale invariant in three dimensions.
Skyrme's proposal was the inclusion of a higher fourth-order term with opposing scaling behaviour to provide the soliton with a scale.

Remarkably, it was shown by Adkins and Nappi \cite{Nappi_1984} that the inclusion of the $\omega$ meson to the NL$\sigma$M alone stabilises the solitons, without the need for the Skyrme term.
This is achieved by considering $\omega$ as a gauge particle associated to $\U(1)_V$ and defining a minimally broken $\U(1)_V$ Lagrangian for spin-1 mesons \cite{Meissner_1986}, with explicit breaking of the gauge invariance by introducing a mass term. 
The abelian nature of the $\omega$-meson means it couples anomalously through the gauged Wess--Zumino (WZ) term \cite{Kaymakcalan_1984}.

In this letter, we study the lower dimensional analogue of Adkins and Nappi's $(3+1)$-dimensional $\omega$-Skyrme model \cite{Nappi_1984}: a variant of the $(2+1)$-dimensional baby Skyrme model, wherein the Skyrme term is removed and the $\omega$-meson is included.
This baby $\omega$-Skyrme model was first studied by Foster and Sutcliffe \cite{Foster_2009}, in which they obtained soliton solutions remarkably similar to those in the baby Skyrme model.
Existence of such solutions was later proven by Greco \cite{Greco_2023}.

We are particularly interested in finding solitons that are crystalline in nature.
The crystalline structure of solitons in the baby Skyrme model was investigated by the author \cite{Leask_2022}, wherein the problem of computing the minimal energy lattice reduced nicely to a simple eigenvalue problem.
Such a method is not possible in the baby $\omega$-Skyrme model.
So we appeal to a method recently constructed for determining crystals in the $(3+1)$-dimensional $\omega$-Skyrme model \cite{Leask_Speight_2024}.
Employing this method enables us to determine the ground state crystalline configuration in the $(2+1)$-dimensional analogue baby $\omega$-Skyrme model.


\section{The baby Skyrme model coupled to the $\omega$-meson}

The baby $\omega$-Skyrme model consists of a single scalar field $\phi: \mathbb{R}\times\mathbb{R}^2 \rightarrow S^2$ coupled to the $\omega$ vector meson, where $(S^2,h,\Omega)$ is the $2$-sphere embedded in $\mathbb{R}^3$ with the induced flat Euclidean metric $h$ and area $2$-form $\Omega$.

In this letter, our main aim is to determine static crystalline solutions within this model.
Therefore, we define $\phi=\varphi \circ \pi$ where $\varphi: \mathbb{R}^2 \rightarrow S^2$ is a fixed map and $\pi: \mathbb{R} \times \mathbb{R}^2 \rightarrow \mathbb{R}^2$ is a projection.
We now identify the baby Skyrme field as the map $\varphi: \mathbb{R}^2 \rightarrow S^2$, which we will normally express as a three vector $\varphi=(\varphi_1,\varphi_2,\varphi_3)$ subject to the unitary condition $\varphi_a\varphi_a = 1$.
In particular, we will study baby Skyrme fields $\varphi:\mathbb{R}^2\rightarrow S^2$ that are periodic with respect to some $2$-dimensional period lattice
\begin{equation}
    \Lambda = \left\{ n_1 \vec{X}_1 + n_2 \vec{X}_2: n_i \in \mathbb{Z} \right\},
\end{equation}
i.e. we impose the condition $\varphi(x+X)=\varphi(x)$ for all $x\in\mathbb{R}^2$ and $X \in \Lambda$.
This is achieved by interpreting the domain of the fields $\varphi,\omega$ as $\mathbb{R}^2/\Lambda$, where $(\mathbb{R}^2/\Lambda,d)$ is a $2$-torus equipped with the standard Euclidean metric $d$.
Following Harland \textit{et al.} \cite{Leask_2023}, we identify this domain with the unit $2$-torus by $\mathbb{T}^2\equiv S^1 \times S^1 = \mathbb{R}^2/\mathbb{Z}^2$ via the diffeomorphism
\begin{equation}
    F: \mathbb{T}^2 \rightarrow \mathbb{R}^2/\Lambda, \quad (x_1,x_2) \mapsto x_1\vec{X}_1 + x_2\vec{X}_2.
\end{equation}
The Euclidean metric $d$ on $\mathbb{R}^2/\Lambda$ can be identified with the pullback metric $g$ on $\mathbb{T}^2$, i.e.\
\begin{equation}
    g = F^*d = g_{ij} \textup{d}x_i \textup{d}x_j, \quad g_{ij} = \vec{X}_i \cdot \vec{X}_j.
\end{equation}
Varying the flat metric $g_s$ on $\mathbb{T}^2$ with $g_0=F^* d$ is equivalent to varying the lattice $\Lambda_s$ with $\Lambda_0=\Lambda$.
Then the energy minimized over variations $g_s$ of the domain metric is equivalent to determining the energy minimizing period lattice $\Lambda_\diamond$.

Since the domain $\mathbb{T}^2$ is compact, the baby Skyrme map $\varphi: \mathbb{T}^2 \rightarrow S^2$ has an associated topological degree given by
\begin{equation}
    B[\varphi] = \int_{\mathbb{T}^2} \textup{d}^2x \sqrt{g} \mathcal{B}^0 \in \mathbb{Z},
\end{equation}
where the conserved topological current is
\begin{equation}
    \mathcal{B}^\mu = -\frac{1}{8\pi\sqrt{g}} \epsilon^{\mu\alpha\beta} \epsilon^{ijk} \varphi_i \partial_\alpha \varphi_j \partial_\beta \varphi_k.
\end{equation}

Further, as we are only concerned with static energy minimizers within the theory, the spatial components of the $\omega$-meson vanish $\omega_i=0$, since the topological current $B^\mu$ acts as a source term for $\omega_\mu$ \cite{Foster_2009}.
So, we opt to drop the subscript and denote $\omega \equiv \omega_0$.
Then, with the above conventions, the static energy functional of this model is defined by
\begin{equation}
\begin{split}
    E = \int_{\mathbb{T}^2} \textup{d}^2x \sqrt{g} \, \left\{ m^2 (1 - \varphi_3) + \frac{1}{2} g^{ij} \partial_i \varphi_a \partial_j \varphi_a \right. \\
    \left. - \frac{1}{2} g^{ij} \partial_i \omega \partial_j \omega - \frac{1}{2} M^2 \omega^2 - c_{\omega} \omega \mathcal{B}_0 \right\},
    \end{split}
\end{equation}
which is not bounded below and renders usual energy minimization methods useless.
The corresponding field equations are found to be
\begin{equation}
\label{eq: Field EoM}
    \Phi_a = -\frac{1}{2} g^{ij} \partial_i\partial_j\varphi_a - m^2 \delta^{a3} + \frac{c_\omega\epsilon^{ij}\epsilon^{abc}}{4\pi\sqrt{g}} \varphi_b\partial_i\omega\partial_j\varphi_c.
\end{equation}
and
\begin{equation}
\label{eq: Omega EoM}
    \left( -g^{ij}\partial_i\partial_j + M^2 \right) \omega = -c_\omega \mathcal{B}_0,
\end{equation}
where $\Phi$ is the tension field of $\varphi$.
It can be seen from Eq.~\eqref{eq: Omega EoM} that the $\omega$-meson is completely determined by the baby Skyrme field $\varphi$ and the domain metric $g$.
Following Gudnason and Speight \cite{Gudnason_2020}, multiplying the $\omega$-meson field equation \eqref{eq: Omega EoM} by $\omega$ and then integrating by parts allows the energy to be rewritten in a more convenient form,
\begin{equation}
\label{eq: Main energy}
    \begin{split}
        E = \int_{\mathbb{T}^2} \textup{d}^2x \sqrt{g} \, \left\{ m^2 (1 - \varphi_3) + \frac{1}{2} g^{ij} \partial_i \varphi_a \partial_j \varphi_a \right. \\
        \left. + \frac{1}{2} g^{ij} \partial_i \omega \partial_j \omega + \frac{1}{2} M^2 \omega^2 \right\},
    \end{split}
\end{equation}
which is bounded below by zero.

The general method for determining crystalline solitons in Skyrme models requires minimizing the static energy functional with respect to variations of the Skyrme field $\varphi$ and the domain metric $g$.
We choose to do this numerically using an accelerated gradient descent based method known as arrested Newton flow.
Formally, we are solving Newton's equation of motion for the potential energy $E$, that is
\begin{equation}
\label{eq: Newton flow}
    \frac{\textup{d}^2}{\textup{d}t^2}(\varphi_a,g_{ij}) = -\nabla E.
\end{equation}
The gradient can be understood by using the calculus of variations.
Let us write $\nabla E=(\Phi_a,S_{ij})$, where $\Phi_a$ and $S_{ij}$ are defined by
\begin{equation}
    \begin{split}
        \frac{\textup{d}}{\textup{d}s}E(\varphi_s,g_s)\bigg|_{s=0} = \int_{\mathbb{T}^2} \textup{d}^2x \sqrt{g} \,\left\{  \Phi_a(\varphi,g)\dot{\varphi}_a \right. \\ \left. + S_{ij}(\varphi,g)\dot{g}_{kl}g^{jk}g^{li}\right\}
    \end{split}
\end{equation}
for all smooth one-parameter variations $\varphi_s, g_s$ with initial conditions $(\varphi_0,g_0)=(\varphi,F^*d)$ and $(\dot{\varphi},\dot{g})=(0,0)$.
The tension field $\Phi$ has already been computed in \eqref{eq: Field EoM} and $S=S_{ij}\textup{d}x_i\textup{d}x_j$ is known as the stress-energy tensor.

The stress-energy tensor $S(\varphi,g) \in \Gamma(\odot^2 T^*\mathbb{T}^2)$ is a symmetric $2$-covariant tensor field on $\mathbb{T}^2$.
This was computed by J\"aykk\"a \textit{et al.} \cite{Jaykka_2012} in the context of the baby Skyrme model.
However, the inclusion of the $\omega$-meson makes this calculation much more difficult, due to $\omega$ depending on $\varphi$ and $g$ via the constraint \eqref{eq: Omega EoM}.
So, the stress-energy tensor is more delicate and requires some further thought.
Nevertheless, the stress-energy tensor was recently computed in \cite{Leask_Speight_2024} for the three dimensional $\omega$-Skyrme model.
Using the methodology laid out therein, the stress-energy tensor associated to the energy functional \eqref{eq: Main energy}, and subject to the $\omega$-meson constraint \eqref{eq: Omega EoM}, is found to be
\begin{equation}
\label{eq: Stress tensor}
    \begin{split}
        S = \left( \frac{1}{4}|\textup{d}\varphi|^2_g + \frac{1}{2}(V\circ\varphi) -  \frac{1}{4} |\textup{d}\omega|^2_g -\frac{1}{4} M^2 \omega^2 \right) g \\ - \left(\frac{1}{2} \varphi^*h - \frac{1}{2} \textup{d}\omega\otimes\textup{d}\omega \right).
    \end{split}
\end{equation}
In a local coordinate system this reads
\begin{align}
\label{eq: Stress tensor local}
    S_{ij} = \, & \left\{ \frac{1}{4}g^{kl} \partial_k\varphi_a\partial_l\varphi_a + \frac{1}{2}m^2\left(1-\varphi_3\right) - \frac{1}{4}g^{kl}\partial_k\omega\partial_l\omega \right. \nonumber \\
    & \left. - \frac{1}{4}M^2\omega^2 \right\} g_{ij} 
    - \frac{1}{2} \partial_i\varphi_a\partial_j\varphi_a + \frac{1}{2} \partial_i\omega\partial_j\omega.
\end{align}
We note that this form for the stress tensor also holds in general for maps $\varphi:(M,g)\to(N,h)$ between Riemannian 2-manifolds.

In order to do numerics, we follow \cite{Leask_2024} and define the metric independent integrals
\begin{align}
\label{eq: MII1}
    V^{\pm} = \, & \int_{\mathbb{T}^2} \textup{d}^2x \, \left( m^2\left(1-\varphi_3\right) \pm \frac{1}{2}\omega^2 \right), \\
\label{eq: MII2}
    L_{ij}^{\pm} = \, & \int_{\mathbb{T}^2} \textup{d}^2x \, \left( \frac{1}{2} \partial_i \varphi_a \partial_j \varphi_a \pm \frac{1}{2} \partial_i \omega \partial_j \omega \right).
\end{align}
Then the energy functional can be written simply as
\begin{align}
    E(\varphi,g) = \, & \sqrt{g} \, g^{ij} L_{ij}^{+} + \sqrt{g}\,V^{+}.
\end{align}
Likewise, the energy gradient with respect to the metric is defined in terms of the metric independent integrals \eqref{eq: MII1} and \eqref{eq: MII2} as
\begin{align}
    \frac{\partial E}{\partial g_{ij}} = \, & \int_{\mathbb{T}^2} \textup{d}^2x \sqrt{g} \, S^{ij} \nonumber \\
    = \, & \frac{1}{2} \sqrt{g} \, g^{ij} V^{-} + \sqrt{g} \left( \frac{1}{2}g^{kl}g^{ij} - g^{ik}g^{jl} \right) L_{kl}^{-},
\end{align}
where the contravariant components of the stress-energy tensor are defined by $S^{ij}=g^{ik}S_{kl}g^{lj}$, and the stress tensor components are given in \eqref{eq: Stress tensor local}.


\section{Baby $\omega$-skyrmion crystals}

\begin{table}[t]
    \centering
    \begin{tabular}{c|cccc}
        \hline
        \hline
        Model & $E$ & $E/(4\pi B)$ & $L$ & $\theta$ \\
        \hline
        Baby $\omega$-Skyrme & $36.013$ & $1.4330$ & $9.59$ & $\pi/3$ \\
        Baby Skyrme & $36.548$ & $1.4543$ & $9.60$ & $\pi/3$ \\
        \hline
        \hline
    \end{tabular}
    \caption{Comparison of the minimal energy $B=2$ hexagonal soliton crystals in the baby $\omega$-Skyrme and baby Skyrme models.}
    \label{tab: Comparison}
\end{table}

To determine soliton crystals within the baby $\omega$-Skyrme model, we need initial configurations for $\varphi,\omega$ and $g$.
For the baby Skyrme field, we consider the axially symmetric configuration in polar coordinates
\begin{equation}
    \varphi_0= \left( \sin f(r) \cos B\theta, \sin f(r) \sin B\theta, \cos f(r) \right),
\label{eq: Initial configurations - Radial ansatz}
\end{equation}
where $f(r)$ is some monotonically decreasing profile function that satisfies the boundary conditions $f(0)=\pi$ and $f(\infty)=0$.
As we only require an approximation for the initial field configuration, we choose the profile function given by \cite{Karliner_2008}
\begin{equation}
    f(r) = \pi \exp(-r).
\label{eq: Initial configurations - Profile function}
\end{equation}

Our initial field configuration $\varphi_0$ is that of the axially symmetric ansatz \eqref{eq: Initial configurations - Radial ansatz} with $B=2$ and the profile function \eqref{eq: Initial configurations - Profile function}.
The initial metric on $\mathbb{T}^2$ is chosen to be $(g_0)_{ij}=L^2\delta_{ij}$, corresponding to some square lattice $\Lambda$ of side length $L$.
A good initial approximation for the $\omega$-meson can be obtained by setting the Laplacian to zero in \eqref{eq: Omega EoM}, which gives $\omega_0=-c_\omega\mathcal{B}_0/M^2$.

With the initial configuration $(\varphi_0,\omega_0,g_0)$ in place, we then apply the arrested Newton flow algorithm detailed above.
If at any time the energy begins to increase, the flow is arrested at that position and the velocities $\frac{\textup{d}}{\textup{d}t}(\varphi_a,g_{ij})$ are set to zero.
The algorithm has deemed to have converged once $\nabla E$ is below some sufficiently small tolerance.
As the $\omega$-meson is dependent upon the field $\varphi$ and metric $g$ via the constraint \eqref{eq: Omega EoM}, computation of the gradient $\nabla E$ requires computing the $\omega$ field at each time step.
Following \cite{Gudnason_2020,Leask_Speight_2024}, we apply a conjugate gradient method to solve the constraint \eqref{eq: Omega EoM} each time step.

For convenience, we select the same constants as Foster and Sutcliffe \cite{Foster_2009}, which were chosen to yield the same $B=1$ soliton energy as the baby Skyrme model.
These are $m=1/\sqrt{10}$, $\kappa=1$ and $c_\omega=20.83$, with the meson mass determined by the relation $c_\omega=4\pi\kappa M$.

\begin{figure}[t]
	\centering
	\begin{subfigure}{0.40\textwidth}
	\includegraphics[width=\textwidth]{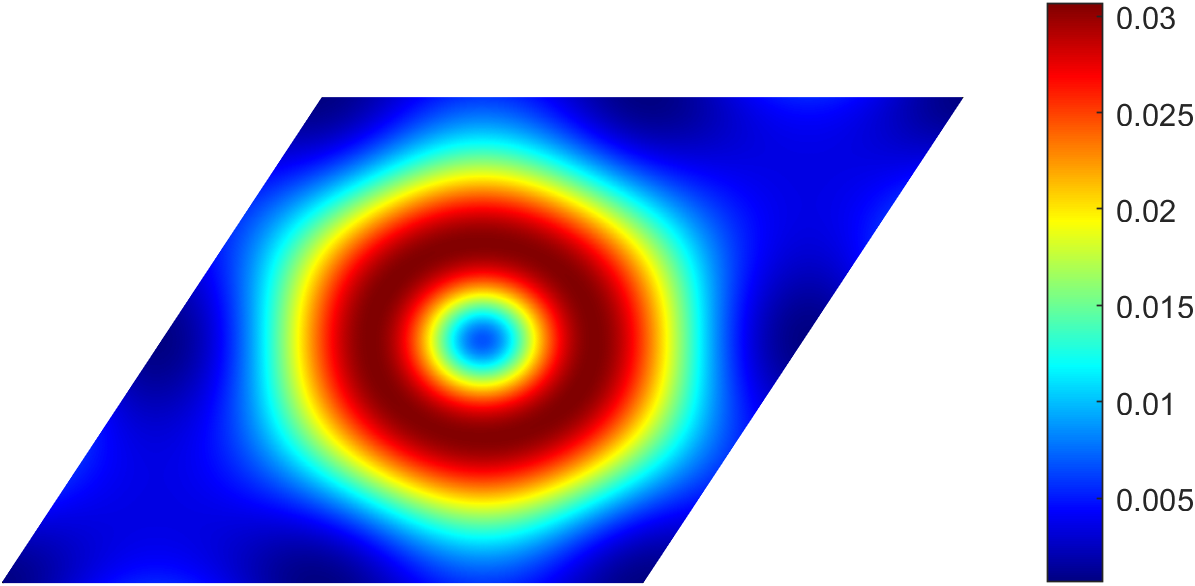}
	\caption{$\mathcal{E}(\vec{x})$}
	\label{fig: Energy density}
	\end{subfigure}
    \\
    \begin{subfigure}{0.40\textwidth}
	\includegraphics[width=\textwidth]{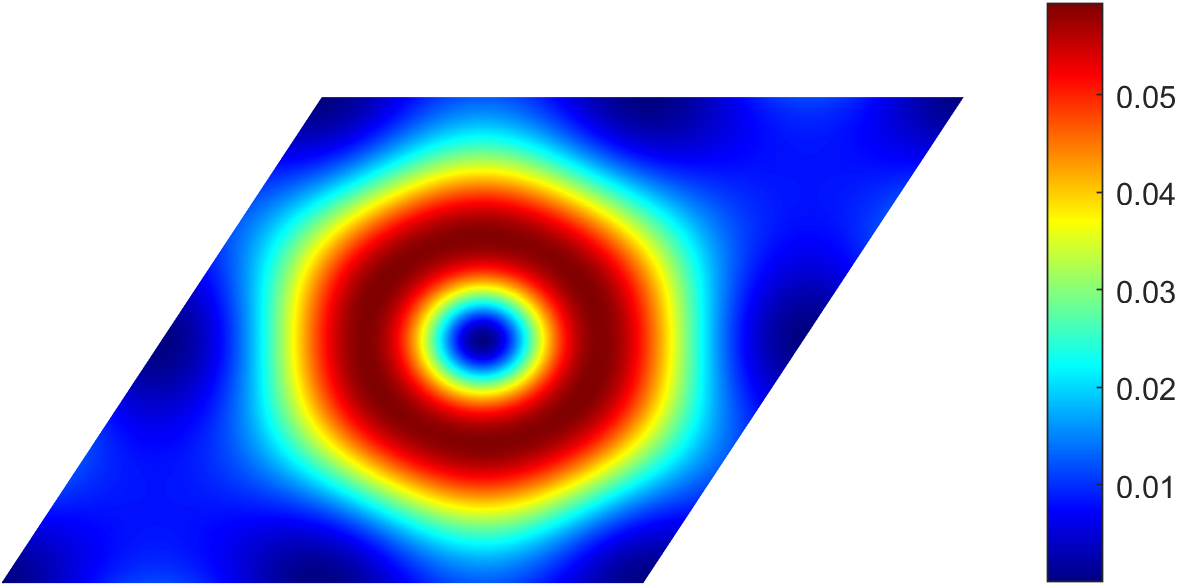}
	\caption{$\mathcal{B}^0(\vec{x})$}
	\label{fig: Charge density}
	\end{subfigure}
    \\
	\begin{subfigure}{0.40\textwidth}
	\includegraphics[width=\textwidth]{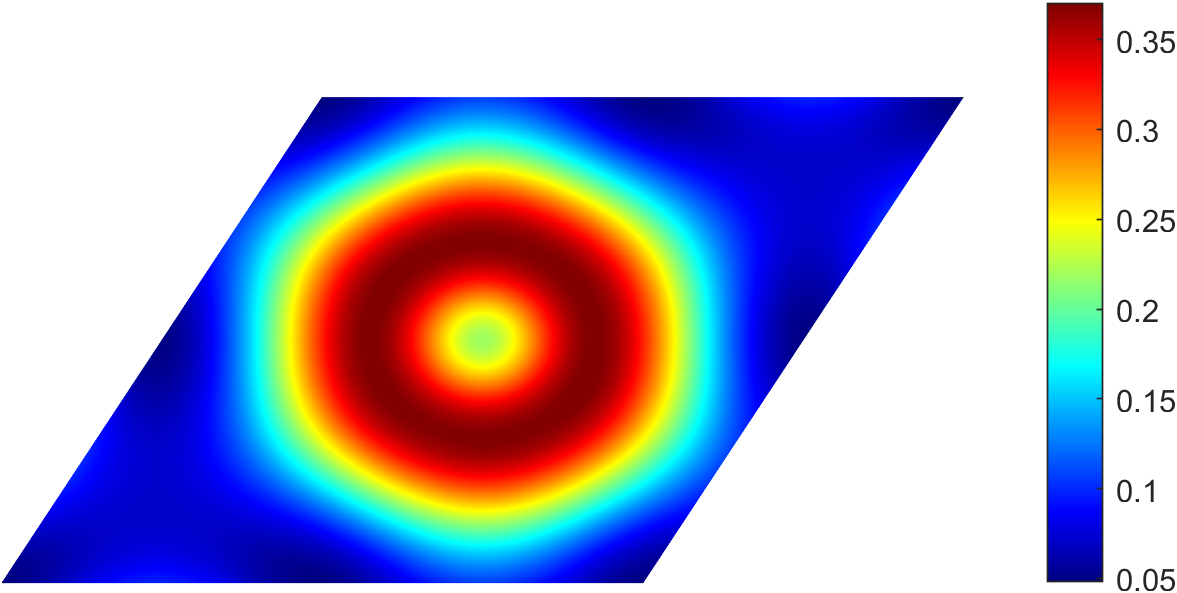}
	\caption{$-\omega(\vec{x})$}
	\label{fig: Omega density}
	\end{subfigure}
	\caption{Density plots of (a) the energy, (b) the charge and (c) the $\omega$-meson for the energy minimizing $B=2$ equianharmonic crystalline solution.}
	\label{fig: Densities}
\end{figure}

The resulting soliton crystal is detailed in Table~\ref{tab: Comparison}, with a comparison to the crystal in the baby Skyrme model with the same parameters.
In both cases, the lattice is equianharmonic with side length $|\vec{X}_1|=|\vec{X}_2|=L$ and the angle between the two lattice vectors is defined by $\cos(\theta)=(\vec{X}_1 \cdot \vec{X}_2)/L^2$.
As was observed in \cite{Foster_2009} for solitons on $\mathbb{R}^2$, the soliton crystal found here in the baby $\omega$-Skyrme model is both qualitatively and quantitatively similar to that of the baby skyrmion crystal of \cite{Leask_2022}.
The resulting $B=2$ hexagonal crystal is plotted in Figure~\ref{fig: Densities}.
Akin to the $(3+1)$-dimensional $\omega$-Skyrme model, the $\omega$-meson here appears to be a smoothed-out version of the charge density $\mathcal{B}_0$ \cite{Manton_2022}.


\section{Concluding remarks}

In this letter, we have presented a numerical method to determine soliton crystals in the $(2+1)$-dimensional baby $\omega$-Skyrme model.
The method detailed herein employs the recently developed algorithm \cite{Leask_Speight_2024} in the higher $(3+1)$-dimensional $\omega$-meson variant of the Skyrme model.
To obtain soliton crystals, we exploited the interpretation of the gradient of the energy with respect to the metric $g$ as the stress tensor $S$ of the field.
The resulting ground state crystal is found to be hexagonal with unit cell charge $B=2$.
This crystal is similar in nature to the hexagonal baby skyrmion crystal observed in \cite{Leask_2022}.

The ground state crystalline configuration in the $(3+1)$-dimensional $\omega$-Skyrme model is dependent on the choice of free parameters of the model \cite{Leask_Speight_2024}.
This is in part due to there being more than one local energy minimizer in this theory.
However, for the baby $\omega$-Skyrme model with the standard pion mass potential, there is currently only one crystal solution.
In the $\omega$-Skyrme model there is a change in the energy ordering of the crystals as the free parameters are varied.
This change in ground state does not happen in the massive and generalized Skyrme models.
To see if a similar effect happens in the baby $\omega$-Skyrme model, a potential that yields multiple crystals would need to be considered, such as the easy plane \cite{Jaykka_2010} or the broken \cite{Jaykka_2012} potentials.

A further natural continuation of this project would be the inclusion of other lower dimensional analogues of vector mesons, such as the $\rho$ meson.
Investigation of such models could provide insight into developing more robust methods for finding crystals coupled to $\omega$ and $\rho$ mesons in higher dimensional theories, such as the so-called HLS model \cite{Yong-Liang_2013}.
The study of crystals in these models are essential for determining an equation of state, which can be used to model dense nuclear matter such as neutron stars \cite{Huidobro_2023}.


\section*{Acknowledgments}

I am grateful to Andrzej Wereszczynski for reading a draft of this manuscript.
This work is supported by a Ph.D. studentship from UKRI, Grant No. EP/V520081/1.


\appendix


\bibliography{bib}

\end{document}